\documentclass[preprint,onecolumn,nofootinbib]{revtex4}

\usepackage[colorlinks=true, linkcolor=blue, urlcolor=blue, citecolor=red, filecolor=black,
            pdfstartview=FitV, pdftitle={}, pdfsubject={}, pdfkeywords={}, 
            pdfpagemode=None, bookmarksopen=true]{hyperref}
\usepackage{graphicx}        
\usepackage{amsmath}         
\usepackage{amssymb}         
\usepackage{amsfonts}        
\usepackage{extarrows}       
\usepackage{color}           
\usepackage{xcolor}          
\usepackage{tikz}            
\usepackage{dcolumn}         
\usepackage{enumerate}       
\usepackage{subcaption}      
\usepackage{caption}         
\usepackage[capitalize]{cleveref} 
\usepackage{ragged2e}        

\begin{document}
\title{
	Spontaneous Vectorization in the Einstein-Maxwell-Vector Model
}
\author{Guang-Zai Ye $^{1}$}
\email{photony@stu2022.jnu.edu.cn}
\author{Chong-Ye Chen $^{1}$}
\email{cychen@stu2022.jnu.edu.cn}
\author{GuoYang Fu $^{2}$}
\email{fuguoyangedu@sjtu.edu.cn}
\author{Chao Niu $^{1}$}
\email{niuchaophy@gmail.com}
\author{Cheng-Yong Zhang $^{1}$}
\email{zhangcy@email.jnu.edu.cn}
\author{Peng Liu $^{1}$}
\email{phylp@email.jnu.edu.cn}
\thanks{Corresponding author}

\affiliation{
	$^1$ Department of Physics and Siyuan Laboratory, Jinan University, Guangzhou 510632, China \\
	$^2$ Department of Physics and Astronomy, Shanghai Jiao Tong University,  Shanghai 200240, China
}

\begin{abstract}

	We investigate spontaneous vectorization in the Einstein-Maxwell-Vector (EMV) model, introducing a novel mechanism driven by the interplay between electromagnetic and vector fields. A key innovation in our work is the resolution of an apparent divergence in the vector field near the event horizon, achieved by employing a generalized coordinate transformation. This not only extends the domain of existence for vectorized Reissner-Nordström black holes (VRNBHs), but also refines the theoretical understanding of such solutions. We introduce a new concept of combined charge $\sqrt{\tilde{Q}^2 + \tilde{P}^2}$, which better captures the underlying physics of these black holes and provides a unified framework for analyzing thermodynamics and observable phenomena such as light ring structures. Our findings suggest that VRNBHs exhibit enhanced thermodynamic preference and distinctive light ring properties compared to Reissner-Nordström solutions. Moreover, we demonstrate how this combined charge approach reveals connections to two-charge black hole solutions, offering promising avenues for observational verification within the context of effective field theories.

\end{abstract}
\maketitle
\tableofcontents

\section{Introduction}

The study of black holes has been a cornerstone of general relativity since its inception. One of the most intriguing results of general relativity is the ``no-hair theorem" \cite{Ruffini:1971bza,Herdeiro:2015waa}. This theorem posits that black holes are remarkably simple objects that can be entirely characterized by three observable parameters: mass ($M$), angular momentum ($J$), and electric charge ($Q$). These parameters define what is called a Kerr-Newman black hole (KNBH) \cite{Newman:1965my}. This theorem implies that all other information about the matter that formed a black hole, referred to as ``hair"--is lost, rendering all black holes with the same values of $M$, $J$, and $Q$ indistinguishable.

However, since the pioneering work of Bartnik and McKinnon \cite{Bartnik:1988am}, who discovered the first self-gravitating Yang-Mills soliton, various new black hole solutions have been found that violate the ``no-hair theorem" (see also the review \cite{Herdeiro:2015waa,Volkov:2016ehx}). These solutions introduce additional fields and mechanisms, allowing black holes to exhibit effective ``hair". One effective method to circumvent the no-hair theorem involves coupling with additional fields \cite{Gibbons:1987ps,Gibbons:1982ih,Doneva:2017bvd,Silva:2017uqg,Cunha:2019dwb,Damour:1993hw,Herdeiro:2018wub,Astefanesei:2019pfq}.

Recent studies have unveiled the phenomenon of spontaneous scalarization in charged black holes \cite{Herdeiro:2018wub,Astefanesei:2019pfq,Myung:2018jvi,Fernandes:2019rez,Blazquez-Salcedo:2020nhs,LuisBlazquez-Salcedo:2020rqp,Hod:2019ulh}. This process demonstrates how the non-minimal coupling between scalar and electromagnetic fields can induce tachyonic instabilities, resulting in the spontaneous formation of scalar ``hair" in Reissner-Nordström black holes (RNBHs). This mechanism is framed within the Einstein-Maxwell-Scalar (EMS) theory. The discovery within this framework prompts the question: Can similar mechanisms induce spontaneous ``hair" formation in higher-order tensor fields, such as vector fields?

In recent years, research on vectorized black holes has garnered significant attention, focusing on areas such as Proca field theories, vector-tensor theories, and so on \cite{Oliveira:2020dru, Herdeiro:2016tmi, Fan:2016jnz, Annulli:2019fzq,Ramazanoglu:2017xbl,Minamitsuji:2020pak, Mai:2023ggs,Kostelecky:2003fs}. In \cite{Oliveira:2020dru}, the EMS theory was extended to incorporate vector fields, resulting in the development of the Einstein-Maxwell-Vector (EMV) theory. While initial studies have demonstrated the existence of vectorized Reissner-Nordström black holes (VRNBHs) and provided phase diagrams, several crucial aspects remain unexplored or misunderstood. Specifically, the physical mechanisms underlying vectorization, the divergences induced by coordinates, the thermodynamic properties of these solutions, as well as the complex interactions between Maxwell and non-gauge vector fields remain insufficiently understood.

This paper addresses the above existing gaps and offers several significant contributions to the field. First, we present a more comprehensive domain of existence for VRNBHs by resolving coordinate-induced divergences that were previously misinterpreted as physical boundaries. This expanded parameter space challenges some conclusions of earlier studies and provides a foundation for our subsequent analyses. Building on this, we propose a novel mechanism for vectorization, framing it as a competition between electromagnetic and vector fields. Additionally, when treating the vector field as a perturbation, we find that the EMV model closely aligns with the two gauge model \cite{Zhang:2020znl,Zhang:2022hxl} in the absence of coupling terms. Basing on this insight, we propose the combined charge \(\sqrt{Q^2+P^2}\) as a fundamental parameter to depict the dynamic interplay between electromagnetic and vector fields. We substantiate this novel mechanism through a comprehensive analysis of thermodynamic properties and light ring structures of VRNBHs. Our comprehensive analysis reveals that the combined charge framework provides a more coherent and natural description of these black holes, shedding light on the complex interplay between electromagnetic and vector fields and their effects on observable properties \cite{EventHorizonTelescope:2019dse}. This unified perspective not only offers a deeper understanding of VRNBHs but also suggests potential observational signatures in effective field theories of gravity.

The structure of this paper is as follows: \cref{sec:2} introduces the fundamental framework of the EMV model, discusses relevant physical quantities, and presents analytical solutions for perturbative backgrounds. In \cref{sec:3}, we delineate the complete domain of existence for VRNBHs and analyze our numerical results. Finally, \cref{sec:4} provides a comprehensive discussion of our findings. Throughout this paper, we employ geometric units where $ G = c = 4\pi \epsilon_0 = 1 $.

\section{The Einstein-Maxwell-Vector Model}\label{sec:2}

\subsection{The Action and Equations of Motion}

We begin by introducing the Einstein-Maxwell-Vector theory, where the electromagnetic field is non-minimally coupled to a real, massless vector field $B^a$ through the coupling function $f(|B|^2)$. The corresponding action is \cite{Oliveira:2020dru}:
\begin{equation}
	\label{eq:action}
	S=\frac{1}{16\pi }\int d^{4}x \sqrt{-g}\Big[R-f(|B|^{2})F^{ab}F_{ab}-V^{ab}V_{ab}\Big],
\end{equation}
where $R$ is Ricci scalar, $F_{ab}=\nabla _{a}A_b-\nabla _{b}A_a$ is the electromagnetic field strength tensor corresponding to the 4-potential $A^a$, and $V_{ab}=\nabla _{a}B_b-\nabla _{b}B_a$ is the vector field strength tensor corresponding to $B^a$. 

Varying the action \cref{eq:action} with respect to the metric yields the field equations:
\begin{equation}
	 R_{ab}-\frac{1}{2}g_{ab}R=2\big(T_{ab}^{V}+T_{ab}^{F}\big)\label{eq:space time}
\end{equation}
where the energy-momentum tensors associated with the vector field and Maxwell field are
\begin{align}
	T_{ab}^{V} \equiv & \; {V_{a}}^{c} V_{bc} - \frac{1}{4} V_{cd} V^{cd} g_{ab}
	+ \frac{1}{2} \frac{df}{d(|B|^{2})} F_{cd} F^{cd} B_{a} B_{b}, \label{eq:emtV}                                          \\[10pt]
	T_{ab}^{F} \equiv & \; f(|B|^{2}) \left( {F_{a}}^{c} F_{bc} - \frac{1}{4} F_{cd} F^{cd} g_{ab} \right). \label{eq:emtF}
\end{align}
The equation of motion for the vector field is given by: 
\begin{equation}
	\nabla _{a}V^{ab}=\frac{1}{2}\frac{df}{d(|B|^{2})}F^{cd}F_{cd}B^{b}\label{eq:vector field}
\end{equation}
and the Maxwell equation is:
\begin{equation}
	\label{eq:gauge field}\nabla_{a}(f\cdot F^{ab})=0 .
\end{equation}

In accordance with \cite{Oliveira:2020dru}, we adopt a quadratic exponential coupling function:
\begin{equation}
	\label{eq:coupling}
	f(|B|^{2}) = \exp(\alpha |B|^{2}).
\end{equation}
Substituting this coupling function into the vector field equation \cref{eq:vector field} yields
\begin{equation}
	\label{eq:vector2}
	\nabla _{a}V^{ab}=\frac{1}{2}\alpha f(|B|^{2})F^{cd}F_{cd}B^{b}={\mu}^{2}_{eff}B^{b}.
\end{equation}
In the case of this work, we have $F^{cd}F_{cd}<0$. Consequently, the effective mass $\mu^2_\text{eff}$ can only be negative when $\alpha > 0$. Under these conditions, the system would exhibit tachyonic instability, potentially leading to spontaneous vectorization.

\subsection{The Ansatz}

Previous work on EMV theory employed Boyer-Lindquist (BL) coordinates, utilizing the following ansatz \cite{Oliveira:2020dru},
\begin{equation}
	\label{eq:BLcoord}
	\begin{aligned}
		ds^{2} & =-\sigma(\bar{r}) ^{2}N(\bar{r})dt^{2}+\frac{d\bar{r}^{2}}{N(\bar{r})}+\bar{r}^{2}d {\Omega _{2}}^{2}
		\\A_{a}&=A_{t}(\bar{r})dt\qquad B_{a}=B_{t}(\bar{r})dt,
	\end{aligned}
\end{equation}
where $\bar{r}$ is the radial coordinates. In \cite{Oliveira:2020dru}, the authors found parameter configurations where the vector field component $B_t$ diverges at the horizon, marking a critical line for solutions. While this might seem to indicate a physical singularity, it is important to note that the scalar quantity $B^2 = B^t B^t g_{tt}$ remains finite. This suggests that no physical singularity occurs at the horizon. The apparent divergence in $B_t$ is, in fact, analogous to the coordinate singularity that appears in the Schwarzschild coordinates for a Schwarzschild black hole. Just as the Schwarzschild coordinate singularity can be resolved through an appropriate coordinate transformation, the divergence in $B_t$ can be addressed similarly. This phenomenon underscores the importance of choosing suitable coordinates when analyzing the behavior of fields near a black hole's event horizon.

This apparent singularity can be effectively eliminated through a generalized coordinate with the following ansatz \cite{Fernandes:2022gde}:
\begin{equation}
	\label{eq:ansatz}
	\begin{aligned}
		ds^{2} & =-h(r)\mathcal{N}(r)^{2}dt ^{2}+\frac{g(r)}{h(r)}(dr^{2}+r^{2}d {\Omega _{2}}^{2}) \\
		A_{a}  & =A_{t}(r)dt\qquad B_{a}=B_{t}(r)dt,
	\end{aligned}
\end{equation}
where $\mathcal{N}(r)=1-r_{H}/r$, and $r_{H}$ is the event horizon location. The functions $h(r)$ and $g(r)$ are radially dependent metric components. The transformation between the BL coordinates and these generalized coordinates is given by:
\begin{equation}\label{eq:coord_trans}
	\bar{r}=r \sqrt{\frac{g(r)}{h(r)}}.
\end{equation}
This coordinate transformation effectively eliminates the coordinate singularity, present in the BL coordinates. The explicit metric of RNBHs in this coordinate system is presented in \cref{sec:AA}. Furthermore, this transformation simplifies the structure of our field equations, facilitating more efficient numerical solutions. A detailed discussion of these simplifications is provided in \cref{sec:NI}.

Having established the metric ansatz for our spherically symmetric spacetime, we now turn our attention to the thermodynamic properties of the system.

\subsection{Thermodynamics}

In asymptotically flat, spherically symmetric spacetimes, the properties of the system can be characterized by a two-dimensional spherical surface, $\partial \Sigma_{\infty}$, defined at constant time $t$ and in the limit $r \rightarrow \infty$ for the radial coordinate. This surface encapsulates the system in its equilibrium state. A key quantity in this equilibrium state is the Arnowitt-Deser-Misner (ADM) mass, which represents the total energy of the spacetime as observed from infinity. For our metric, the ADM mass can be calculated as follows \cite{Baumgarte_Shapiro_2010,Garcia:2023ntf}:
\begin{equation}
	\label{eq:ADMM}
	\begin{aligned}
		M & \equiv \frac{1}{16\pi  }\oint_{\partial \Sigma _{\infty}}\left[\partial^{b}\gamma _{ab}-\partial_{a}(\delta ^{cd}\gamma _{cd})\right]{dS}^{a} \\
		  & =\frac{r^{2}\sqrt{g(r)}\left[g(r)h'(r)-h(r)g'(r)\right]}{2h(r)^{2}\sqrt{h(r)}} \Bigg|_{r\to \infty},
	\end{aligned}
\end{equation}
where $\gamma _{ab}$ is the induced metric on the hypersurface ($\partial \Sigma _{t}$), $\delta _{ab}$ is Euclidean spatial metric, $\partial _{a}$ is the ordinary derivative operator, and $dS^{a}$ is the oriented surface element of $\partial \Sigma_{\infty}$.

Additionally, the ADM mass can also be obtained by calculating the conserved charge associated with the timelike Killing vector field $\xi^{a}=(\partial _{t})^{a}$.
This approach allows us to decompose the mass into contributions from the horizon and the matter fields \cite{Baumgarte_Shapiro_2010,Garcia:2023ntf}:
\begin{equation}
	\label{eq:KormarM}
	M\equiv M_{H}+M_{F}+M_{V},
\end{equation}
where $M_{H}$ represents the horizon mass, while $M_F$ and $M_V$ are the masses associated with the electromagnetic field and vector field outside the horizon, respectively:
\begin{align}
	M_{H} & \equiv - \frac{1}{8\pi } \oint _{H}\nabla ^{a}\xi^{b}d S_{ab}^{H}=r_{H} \sqrt{g(r_{H})},\label{eq:Mh} \\
	M_{F} & \equiv \int_{\partial \Sigma_t}{dS}^{a}(2T_{ab}^{F}\xi^{b}-T^{F}\xi_{a})=Q \Phi, \label{eq:Mf}        \\
	M_{V} & \equiv \int_{\partial \Sigma_t}{dS}^{a}(2T_{ab}^{V}\xi^{b}-T^{V}\xi_{a})=0.\label{eq:Mv}
\end{align}
The vanishing of $M_V$ is proved in \cref{sec:AC}.
The electric charge $Q$ and the vector ``charge'' $P$ can be extracted from the asymptotic behavior of the temporal components of the gauge fields:
\begin{equation}
	A_{t}\sim \Phi-\frac{Q}{r}+ \cdots ,\qquad B_t\sim \frac{P}{r}+ \cdots,
\end{equation}
where $\Phi$ is the electric potential at infinity.
For our system, the Hawking temperature $T_H$ and the Bekenstein-Hawking entropy $S$ are given by:
\begin{equation}\label{eq:TS}
	T_{H}=\frac{h(r_{H})}{2\pi r_{H}\sqrt{g(r_{H})}},\qquad
	S=\frac{\pi r_{H}^{2}g(r_{H} )}{h(r_{H})}.
\end{equation}

The physical quantities mentioned above are related by the Smarr mass formula
\begin{equation}\label{eq:smarr}
	M=2T_{H}S+Q \Phi.
\end{equation}
These quantities are also connected via the first law of black hole thermodynamics:
\begin{equation}
	dM=T_{H}dS+\Phi dQ.
\end{equation}
Finally, observe that \cref{eq:KormarM,eq:Mh,eq:Mf,eq:Mv,eq:smarr} are consistent with a different Smarr relation, which is expressed solely in terms of horizon quantities:
\begin{equation}
	\label{eq:Smarrrh}
	M_H=2 T_H S.
\end{equation}
To study thermodynamic instability of this system, we also consider the free energy:
\begin{equation}
	\label{eq:Feq}
	F=M-TS.
\end{equation}

Following the definition of black hole thermodynamic quantities, we now focus on another crucial aspect of black hole physics: the light ring structure. This feature is not only theoretically significant but also observationally relevant in the era of direct black hole imaging.

\subsection{Light Ring}

The Event Horizon Telescope (EHT) has captured the first-ever image of a black hole's shadow, which is closely related to the light ring structure \cite{EventHorizonTelescope:2019dse,
	EventHorizonTelescope:2019uob,
	EventHorizonTelescope:2019jan,
	EventHorizonTelescope:2019ths,
	EventHorizonTelescope:2019pgp,
	EventHorizonTelescope:2019ggy}. 
This groundbreaking observation opens up new possibilities for studying potential deviations from general relativity, such as the presence of additional fields like scalar or vector fields. The high-precision data from the EHT provides an unprecedented opportunity to explore such deviations, including those involving spontaneous vectorization. By carefully analyzing the properties of the light ring and shadow, we may be able to identify observational signatures of such phenomena\cite{Chen:2022scf}.

Light rings, which are circular null geodesics, play a crucial role in understanding the behavior of photons in strong gravitational fields. To rigorously analyze these phenomena, we consider a photon with a 4-velocity $\left(\frac{\partial}{\partial \lambda}\right)^a$, where $\lambda$ is an affine parameter. The trajectory of such a photon in a given spacetime is governed by the geodesic equation:
\begin{equation}
	\frac{d^2x^\mu}{d\lambda^2} + \Gamma^\mu{_{\alpha\beta}}\frac{dx^\alpha}{d\lambda}\frac{dx^\beta}{d\lambda} = 0,
	\label{eq:PhotonGeodesic}
\end{equation}
where $\Gamma^\mu{_{\alpha\beta}}$ are the Christoffel symbols. For circular orbits, we impose the conditions $\dot{r} = 0$ and $\ddot{r} = 0$, where $\dot{r} \equiv \frac{dr}{d\lambda}$. These conditions, combined with the metric components, lead to the equation for the light ring radius:
\begin{equation}
	\frac{h'(r)}{h(r)} - \frac{g'(r)}{2g(r)} + \frac{1}{r-r_{H}} - \frac{2}{r} = 0.
	\label{eq:LightRingRadiusCondition}
\end{equation}
Here, $h(r)$ and $g(r)$ are functions determined by the metric, and $r_{H}$ is the horizon radius. For comparison, in the case of RNBHs, which describes charged, non-rotating black holes, the light ring radius is given by \cite{Fernandes:2022gde}:
\begin{equation}
	r^{RN}_{LR} = \frac{1}{4} \left(M + \sqrt{9M^2 - 8Q^2} + \sqrt{2M\left(\sqrt{9M^2 - 8Q^2} + 3M\right) - 4Q^2}\right),
	\label{eq:RNLightRingRadius}
\end{equation}
where $M$ is the mass and $Q$ is the charge of the black hole.

To facilitate a more intrinsic description of our system's behavior, we exploit its rescaling symmetry $r \to \lambda r, Q \to \lambda Q, \lambda \in \mathbb{R}^+$. This allows us to introduce the following dimensionless physical quantities:
\begin{equation}
	\begin{aligned}
		\tilde{Q}     & \equiv \frac{Q}{M},  & \tilde{P} & \equiv \frac{P}{M}, & \tilde{S}      & \equiv \frac{S}{4\pi M^{2}}, \\
		\tilde{T}_{H} & \equiv 8\pi M T_{H}, & \tilde{F} & \equiv \frac{F}{M}, & \tilde{r}_{LR} & \equiv \frac{r_{LR}}{M}.
	\end{aligned}
\end{equation}
These dimensionless variables provide a concise and scale-independent description of the system's properties, enabling more effective comparisons between different black hole solutions and potential observational signatures.

With all relevant quantities defined, we will derive the perturbation solution in general coordinates before proceeding to the numerical discussion.

\subsection{Perturbation Solutions}

In examining \cref{eq:vector2}, we identify a tachyonic instability in the vector field under certain configurations.
This instability manifests as the exponential growth of perturbations, indicating the presence of perturbative solutions within the context of RNBHs. These perturbative solutions are essential for defining the existence line for haired black hole, which serves as a boundary delineating the domain of distinct physical states within the system's parameter space.

When the vector field $B^{a}$ is sufficiently small, its back-reaction on the background can be neglected, and the vector field can be treated as a perturbation. By substituting the RN metric (\cref{eq:hgRN,eq:AtRN,eq:rhRN}) into the vector equation (\cref{eq:vector field}), and adopting a linear approximation for the coupling function $f \approx 1 + \alpha B_a B^a$, we obtian:
\begin{equation}
	\label{eq:peomv}
	\begin{aligned}
		B_t''(r) & +4 \left(\frac{2 r}{M^2-Q^2-4 r^2}+\frac{1}{M-Q+2 r}+\frac{1}{M+Q+2 r}\right) B_t'(r)
		\\&+\frac{16 \alpha  Q^2 }{(M-Q+2r)^2 (M+Q+2 r)^2}B_t(r)=0.
	\end{aligned}
\end{equation}
Applying the coordinate transformation
$$
\zeta=\frac{Q^2 \left(-M^2+Q^2+4 r^2\right)^2}{\left(M^2-Q^2\right) \left(M^2+4 M r-Q^2+4 r^2\right)^2}
$$
to \cref{eq:peomv} yields:
\begin{equation}
	\label{eq:hgf}
	(1+\zeta)B_t''(\zeta)+\frac{1}{2}B_t'(\zeta)+\frac{\alpha}{4z} B_t(\zeta)=0.
\end{equation}
This equation admits a solution in terms of hypergeometric functions:
\begin{equation}\label{eq:solpv}
	B_t(\zeta)=\zeta \cdot {}_{2}F_{1}\left(\frac{3}{4}-\frac{1}{4}\sqrt{1-4\alpha},\frac{3}{4}+\frac{1}{4}\sqrt{1-4\alpha   },2,-\zeta\right) .
\end{equation}
In the $\zeta$ coordinate, the horizon and spatial infinity are represented by $ \zeta \in \left[0, \frac{Q^2}{M^2 - Q^2}\right]$.
Notably, \cref{eq:solpv} automatically satisfies the boundary condition for the vector field at the horizon, with real solutions existing only for $\alpha > 1/4$.
To determine the existence line of VRNBHs, we impose the boundary condition $B_t(\zeta)=0$ at $\zeta = \tilde{Q}^{2} / (1-\tilde{Q}^{2})$. Thus, the problem reduces to solving for the charge $\tilde{Q}$:
\begin{equation}\label{eq:psol2}
	\frac{\tilde{Q}^2}{1-\tilde{Q}^2} \cdot {}_{2}F_{1}\left(\frac{3}{4}-\frac{1}{4}\sqrt{1-4\alpha},\frac{3}{4}+\frac{1}{4}\sqrt{1-4\alpha   },2,-\frac{\tilde{Q}^2}{1-\tilde{Q}^2}\right)=0,
\end{equation}
for each $\alpha >1/4$. 
Moreover, we verified that, in the BL coordinate, the existence line equation remains consistent with \cref{eq:psol2} as well\footnotemark[1]\footnotetext[1]{The detailed calculations and analysis for the BL case are available in our open-source repository at \cite{weblink}}.
\begin{figure}[htbp]
	\centering
	\includegraphics[width = 0.5\textwidth]{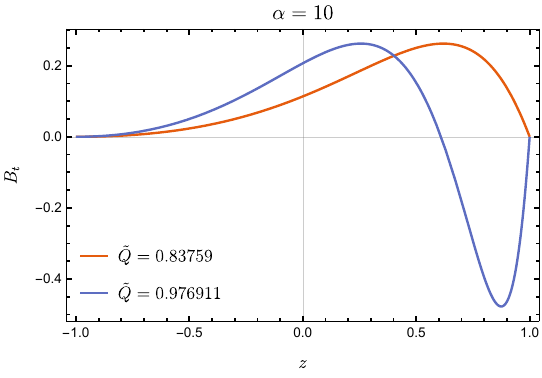}
	\caption{
		\justifying
		There are two perturbative solutions that satisfy the boundary condition $ B_t(r_{H}) = B_t(\infty) = 0 $ at $\alpha = 10$. These correspond to a nodeless solution and a single-node solution, respectively. The $z$-axis is defined by $ z = 1 - \frac{2r_H}{r} $.
	}
	\label{fig:psol}
\end{figure}
For any given $\alpha$, there are multiple values of $\tilde{Q}$ that satisfy the boundary condition. Among these, the smallest $\tilde{Q}$ corresponds to the nodeless configuration. \cref{fig:psol} illustrates the perturbative solutions for nodeless and single-node configurations at $\alpha = 10$. This paper primarily focuses on the behavior of the nodeless configuration, as it is more stable than configurations with nodes.

After examining the full range of values for $\alpha$, we identify the corresponding nodeless values of $\tilde{Q}$, forming an existence line within the $\tilde{Q} - \alpha$ parameter space. In the following section, we conduct a detailed numerical analysis to further investigate the behavior of spontaneous vectorization.

\section{Numerical Solutions and the Mechanism of Vectorization}\label{sec:3}

In this section, we numerically study the domain of existence, thermodynamic properties, the interplay between the vector and electromagnetic fields and light ring characteristics of VRNBHs.

\subsection{Domain of Existence}\label{sec:domain}

In this subsection, we explore the domain of existence for VRNBHs by analyzing the interplay between the electric charge $ \tilde{Q} $ and the vector ``charge" $ \tilde{P} $. Unlike the electric charge, $ \tilde{P} $ is not a globally conserved quantity; however, in the weak field limit where the coupling constant $ \alpha > \frac{1}{4} $, the vector field $ B $ exhibits behavior akin to a gauge field. This similarity motivates us to consider the system in a manner like a two-charge black hole (TCBH) \cite{Zhang:2020znl,Zhang:2022hxl}, where the combined charge $ \tilde{Q}_{com}\equiv \sqrt{\tilde{Q}^2 + \tilde{P}^2} $ serves as a more genuine and insightful parameter for analysis. Next, we present the domain of existence for VRNBHs in the $ \tilde{Q}-\alpha $ and $ \tilde{Q}_{com} - \alpha $ parameter spaces, respectively.

\begin{figure}[htbp]
	\centering
	\subcaptionbox{\label{fig:phaseq}}
	{\includegraphics[width = 0.48\textwidth]{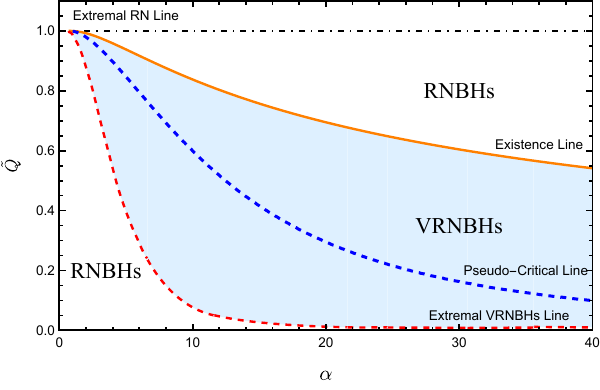}}
	\hfill
	\subcaptionbox{\label{fig:phaseqp}}
	{\includegraphics[width = 0.48\textwidth]{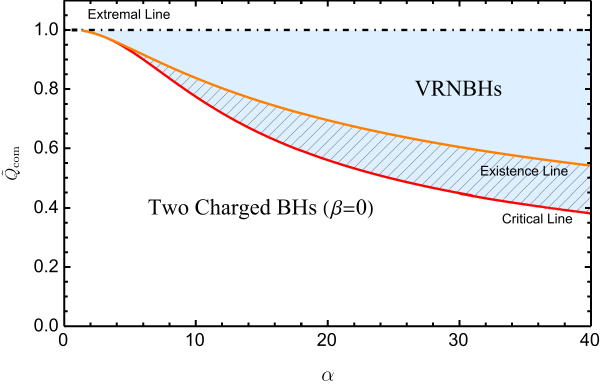}}
	\caption{
		\justifying
		\textbf{(a)}: Domain of existence for VRNBHs (light blue region) in the $\left( \tilde{Q},\alpha \right)$ plane. The white background regions represent RNBHs. The existence line (solid orange line) and the extremal VRNBHs (dashed red line) fall within the domain of RN black holes. The black dotdashed line represents the extremal RN black holes, while the blue dashed line denotes the pseudo-critical line, where the first derivative of $ B_t $ diverges at the horizon within the BL framework. \textbf{(b)}: Domain of existence for VRNBHs (light blue region) in the $\left(\tilde{Q}_{com}, \alpha\right)$ plane, with the white rectangular background region representing TCBHs with $\beta = 0$. The black dotdashed line corresponds to the extremal (zero temperature) line for both VRNBHs and TCBHs. The red line denotes the critical line, while the orange line indicates where $\tilde{P}\thickapprox 0$, corresponds to the existence line in the \cref{fig:phaseq}. Between the existence line and the critical line lies a shaded region where two distinct nodeless solution configurations coexist.
	}\label{fig:phase}
\end{figure}

\cref{fig:phaseq}\footnotemark[2]
\footnotetext[2]{We find that the existence line presented in \cite{Oliveira:2020dru} may be inaccurate. For a comprehensive derivation, please refer to our GitHub repository \cite{weblink}.} presents the complete domain of existence for the nodeless VRNBHs in the $\tilde{Q}-\alpha $ parameter space. The light blue region represents the area where VRNBHs can exist, bounded by the existence line (orange solid line) and the extremal VRNBHs line (red dashed line). The white background regions represent the RNBHs.

A notable feature in \cref{fig:phaseq} is the blue dashed line situated between the existence line and the extremal VRNBHs line. This line, referred to as the pseudo-critical line, corresponds to the critical line within the BL coordinate \cite{Oliveira:2020dru}, where all VRNBHs exhibit a divergence in the derivative of $B_t$ at the horizon. Our analysis indicates that this divergence does not signify a physical spacetime singularity; instead, it results from the choice of coordinates.

To resolve this, we employ a coordinate transformation defined in \cref{eq:coord_trans}, which effectively removes the apparent singularity. This transformation not only clarifies the true boundary of the physical domain but also extends the domain of existence beyond the original coordinate system's limitations.

The generalized coordinate system significantly simplifies the analysis by allowing an analytic solution of the metric function $ g(r) $ (see \cref{eq:g3})\footnotemark[3]. Despite $ \tilde{P} $ and $ \tilde{Q} $ not being parallel, the combination $\tilde{Q}_{com}\equiv \sqrt{\tilde{Q}^{2}+\tilde{P}^{2}} $ emerges naturally in the field expansions (\cref{eq:infexp}), facilitating a more straightforward analysis. 
\footnotetext[3]{For detailed numerical implementations of these solutions, please refer to \cref{sec:NI}, where we also demonstrate the analytical solution for $ g $ within this framework.}

To further elucidate the role of the combined charge $ \tilde{Q}_{com} $, we transition to the discussion within parameter space $ \left(\tilde{Q}_{com}, \alpha \right) $, as depicted in \cref{fig:phaseqp}. This approach aligns with the two gauge fields theory \cite{Zhang:2020znl} and provides a clearer characterization of VRNBHs. In this $ \left(\tilde{Q}_{com}, \alpha \right) $ plane, VRNBHs exist within the light blue region, while the white background corresponds to TCBHs with $\beta = 0$. The red line indicates the critical line between TCBHs and VRNBHs. The black dotdashed line represents the extremal limit where $ \tilde{Q}_{com} = 1 $, marking the boundary beyond which both VRNBHs and TCBHs become extremal.

\begin{figure}[htbp]
	\centering
	\subcaptionbox{\label{fig:phasev3}}
	{\includegraphics[width = 0.48\textwidth]{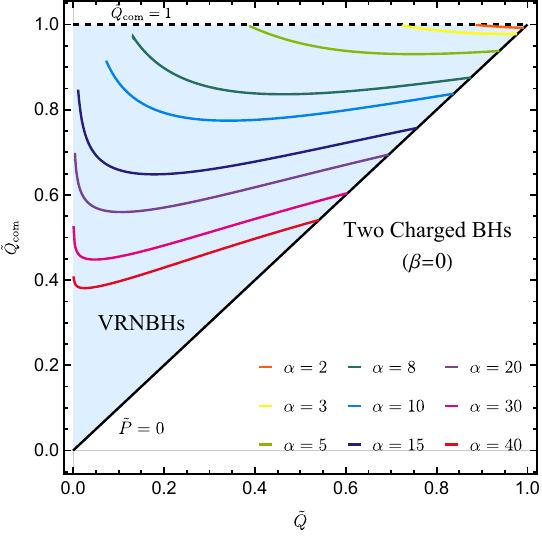}}
	\hfill
	\subcaptionbox{\label{fig:qvsp}}
	{\includegraphics[width = 0.47\textwidth]{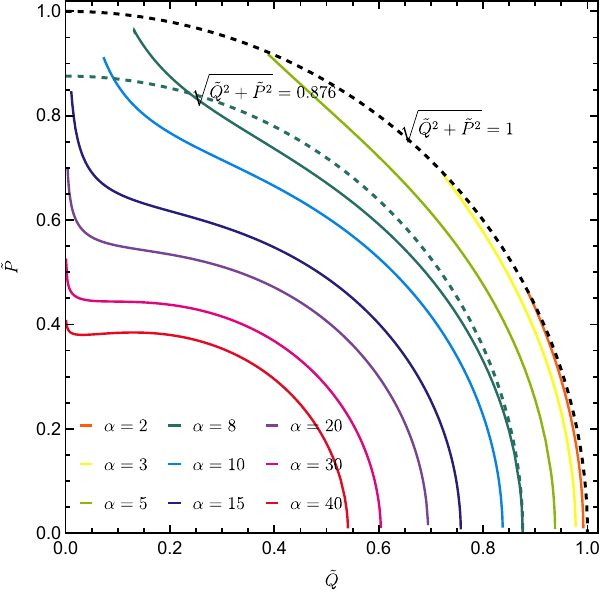}}
	\caption{
		\justifying
		\textbf{(a)}: This plot illustrates the domain of existence for black hole solutions in the $\left(\tilde{Q}, \tilde{Q}_{com}\right)$ plane for various values of the coupling constant $\alpha$ (color-coded). The diagonal line $(\tilde{P}=0)$ represents RNBHs with no vector charge. The horizontal dashed line $\tilde{Q}_{com} = 1$ represents the extremal line where the combined charge is maximized. The area with a white background, which covering the entire parameter space, indicates TCBHs region, and the blue shaded region corresponds to the VRNBHs domain.
		\textbf{(b)}: This plot depicts the relationship between the vector ``charge'' $( \tilde{P} )$ and the electric charge $( \tilde{Q} )$ for  different values of $ \alpha $. The black dashed line $ \sqrt{\tilde{Q}^2 + \tilde{P}^2} = 1 $ represents the extremal limit.The green dashed line represents TCBHs that satisfy $\sqrt{\tilde{Q}^{2}+\tilde{P}^{2}}=0.876$.}
\end{figure}

From the traditional single charge perspective, the domain of existence for VRNBHs shows a solution family that extends from the existence line towards decreasing $Q$ until reaching the extremal line, with each $Q$ value corresponding to a unique solution. However, when viewed through the combined charge framework, the solution family exhibits a different structure: it begins from the critical line and extends in the direction of increasing \(Q_{com}\) towards both the existence line and the extremal line. Within the shaded region of \cref{fig:phaseqp}, a single \(Q_{com}\) value can correspond to two distinct solution configurations (e.g., \(Q_1^2+P_1^2=Q_2^2+P_2^2\) with \(Q_1\neq Q_2\)). This two solution phenomenon is more clearly illustrated in \cref{fig:phasev3}.

\cref{fig:phasev3} illustrates the relationship between $\left(\tilde{Q},\tilde{Q}_{com}\right)$. The black solid line represents the case where $\tilde{P}=0$, corresponding to the existence line in \cref{fig:phase}, while the black dashed line indicates the extremal black holes, given by $\tilde{Q}^{2}+\tilde{P}^{2}=1$. The colored lines depict a series of VRNBHs at various values of $\alpha $. The light blue region and white background region represent the areas corresponding to VRNBHs and TCBHs, respectively. Additionally, for larger values of $\alpha$, the curves become non-monotonic, which explains the origin of the shaded region in \cref{fig:phaseqp}.

The introduction of the combined charge $\tilde{Q}_{com} $ fundamentally reshapes our understanding of the EMV model. Although the vector charge $ \tilde{P} $ is not an independent degree of freedom and is typically determined by the electric charge $ \tilde{Q} $, analyzing the system solely in the $ (\tilde{Q}, \alpha) $ space neglects the nuanced interplay between $ \tilde{P} $ and $ \tilde{Q} $. By adopting $ \tilde{Q}_{com} $ as a primary parameter, we capture the essential dynamics of the system more effectively. This approach not only simplifies the analysis but also reveals new physical phenomena, such as the double solution regions and the intricate structure of light rings, which are pivotal for understanding the thermodynamic stability and spacetime structure of VRNBHs.

\subsection{Competition Between two Fields}

In this subsection, we delve deeper into the competition between two fields in the EMV model, with a focus on a novel mechanism of vectorization from the charge interplay. Specifically, this competition becomes obvious when considering the combined charge $ \tilde{Q}_{com} $, which encapsulates the contributions from both the electromagnetic field (represented by $ \tilde{Q} $) and the vector field (represented by $ \tilde{P} $).

For an asymptotically flat black hole, the $g_{tt}$ component of the metric can be written as:
\begin{equation}
	\label{eq:gttasy}
	g_{tt}= h(r)\mathcal{N}(r)^{2}=1-\frac{2M}{r}+\cdots .
\end{equation}
Expanding the above expression leads to the following form for $h(r)$:
\begin{equation}
	\begin{aligned}
		\label{eq:hasy}
		h(r) & =1-\frac{2(M-r_H)}{r}+\cdots                                                                  \\
		     & \xlongequal{\text{\cref{eq:infexp}}}1-\frac{2\left(\sqrt{P^{2}+Q^{2}+4r_H^{2}}-r_{H}\right)}{r}+\cdots .
	\end{aligned}
\end{equation}
From this, the ADM mass ($M$) can be derived as:
\begin{equation}
	\begin{aligned}
		\label{eq:MeqPQ}
		M & =\sqrt{Q^{2}+P^{2}+4r_H^{2}}                                     \\
		  & \xlongequal{\text{\cref{eq:TS}}}\sqrt{Q^{2}+P^{2}+(2 T_H S)^{2}} \\
		  & \xlongequal{\text{\cref{eq:Smarrrh}}}\sqrt{Q^{2}+P^{2}+M_H^{2}}
	\end{aligned}
\end{equation}
Thus, the black hole mass is determined by a combined charge and horizon properties. The ADM mass is influenced by the Maxwell field and vector fields through their charges, $Q$ and $P$ respectively. This naturally leads to a competitive interaction between the two fields, as both affect the combined charge and energy of the black hole. Furthermore, the dimensionless quantities $ \tilde{Q} $ and $ \tilde{P} $ satisfy the relation:
\begin{equation}
	\label{eq:qpeq}
	\begin{aligned}
		M^{2}                                & =Q^{2}+P^{2}+M_H^{2}                                                            \\
		                                     & \xlongequal[\text{\cref{eq:Smarrrh}}]{\text{\cref{eq:smarr}}}(M_{H}+Q \Phi)^{2} \\
		\implies \tilde{Q}^{2}+\tilde{P}^{2} & =2 \tilde{Q} \Phi \frac{M_H}{M}+\tilde{Q}^{2} \Phi^{2},                         \\
	\end{aligned}
\end{equation}
where $ \Phi $ is the electric potential associated with the charge $ \tilde{Q} $.

As illustrated in \cref{fig:qvsp}, the competition between $ \tilde{Q} $ and $ \tilde{P} $ becomes evident as the coupling constants $ \alpha $ increases. The black dashed line corresponds to extremal black holes, where the combined charge reaches its maximum, while the green dashed line represents a specific baseline case with $\sqrt{\tilde{Q}^2 + \tilde{P}^2} = 0.876$. 
Across various values of $ \alpha $, we observe an inverse correlation between $ \tilde{Q} $ and $ \tilde{P} $, reinforcing the notion of competition between the two fields.

An important result emerges when the charge $\tilde{P}$ associated with the vector field is treated as a perturbation, occurring at either small $\alpha$ or small $\tilde{P}$. Under these conditions, the combined charge relation aligns with that of the two gauge field cases. As depicted in \cref{fig:qvsp}, we chose parameter pair ($\tilde{Q}=0.876, \alpha =8$; For any $\alpha $, the corresponding, unique $\tilde{Q}$ can be deduced from existence line equations \cref{eq:psol2}.) as the example. Here, the green dashed line and green solid line nearly coincide under perturbative conditions.

Analytically, for TCBHs, the relationship between $\tilde{Q}$ and $\tilde{P}$ is given by:
\begin{equation}
	\label{eq:qpeqDG}
	\tilde{Q}^{2}+\tilde{P}^{2}\xlongequal{\text{\cref{eq:tcbhsmarr}}}2\tilde{Q}\Phi\frac{M_H}{M}+\tilde{Q}^{2}\Phi^{2}+2\tilde{P} \Psi \frac{M_H}{M}+\tilde{P}^{2} \Psi^{2}+2\tilde{Q} \Phi \tilde{P} \Psi  ,\qquad\text{for TCBHs}
\end{equation}
where $\Phi,\Psi$ are the gauge field potentials associated with charges $\tilde{Q}$ and $\tilde{P}$, respectively. When the gauge field associated with charge $\tilde{P}$ is treated as a perturbation, the relation in \cref{eq:qpeqDG} approximates that in \cref{eq:qpeq}, noting that $\tilde{P} \Psi$ is on the order of $\mathcal{O}(\tilde{P}^{2})$.
This suggests that, in the perturbative regime, the system is very close to TCBHs. The parameter $\alpha$ measures the deviations of the system from TCBHs.

However, as the coupling constant $ \alpha $ increases, the competition between $ \tilde{Q} $ and $ \tilde{P} $ becomes more pronounced. The vectorization mechanism, therefore, can be understood as a shift in dominance between the electromagnetic and vector fields, governed by the coupling constant $ \alpha $. Larger values of $ \alpha $ amplify the deviations from the two charge black hole regime, leading to significant modifications in the black hole's properties due to the vector field.

As we approach extreme states (i.e., zero-temperature conditions where $ M_H \to 0 $), the combined charge relation simplifies. For VRNBHs, the relation becomes:
$$
\tilde{Q}^2 + \tilde{P}^2 = \tilde{Q}^2 \Phi^2 = \frac{M^2}{M^2} = 1,
$$
while for TCBHs, the relation remains unchanged. Implementing the zero-temperature conditions into \cref{eq:qpeqDG} gives:
$$
\tilde{Q}^2 + \tilde{P}^2 = \tilde{Q}^2 \Phi^2 + \tilde{P}^2 \Psi^2 + 2 \tilde{Q} \Phi \tilde{P} \Psi = \frac{M^2}{M^2} = 1.
$$
This analytical result reinforces the viewpoint that in extremal cases, the contributions from the vector and electromagnetic fields combine to provide a unified description of the black hole's charge.

The relation $\sqrt{\tilde{Q}^2+\tilde{P}^2}=1$ for extreme VRNBHs arises directly from our generalized coordinates introduced in \cref{eq:ansatz}. This coordinate choice allows for a more natural description of the extreme limit. Unlike previous studies using BL coordinates, which could not reach the true extreme state due to obstruction by the pseudo-critical boundary, our approach provides a clearer picture of the behavior of extreme VRNBHs. This underscores the importance of appropriate coordinate choices in studying black holes.

\begin{figure}[htbp]
	\centering
	\subcaptionbox{\label{fig:sq}}
	{\includegraphics[width = 0.46\textwidth]{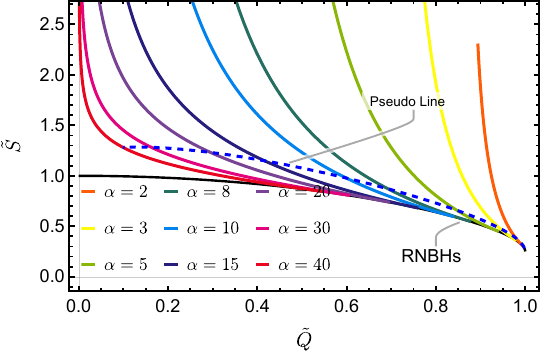}}
	\hfill
	\subcaptionbox{\label{fig:Fq}}
	{\includegraphics[width = 0.46\textwidth]{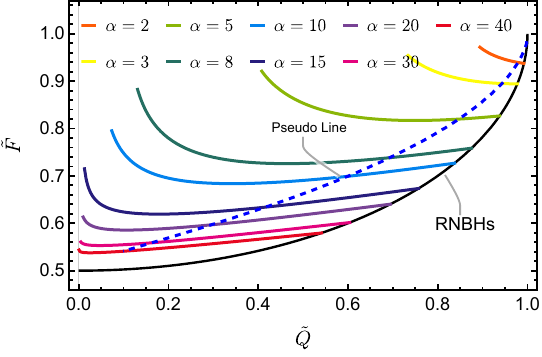}}
	\caption{
		\justifying
		\textbf{(a)}: Reduced entropy $\tilde{S}$ vs $\tilde{Q}$. \textbf{(b)}: Reduced free energy $\tilde{F}$ vs $\tilde{Q}$. The black line represents the RNBHs, while the blue dotted line denotes the pseudo-critical line. The colored solid lines represent VRNBHs at different $\alpha$ values. } \label{fig:qsF}
\end{figure}

\begin{figure}[htbp]
	\centering
	\subcaptionbox{\label{fig:sqp}}
	{\includegraphics[width = 0.48\textwidth]{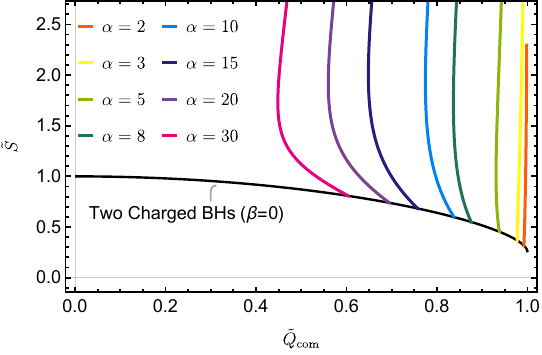}}
	\hfill
	\subcaptionbox{\label{fig:freeqp}}
	{\includegraphics[width = 0.48\textwidth]{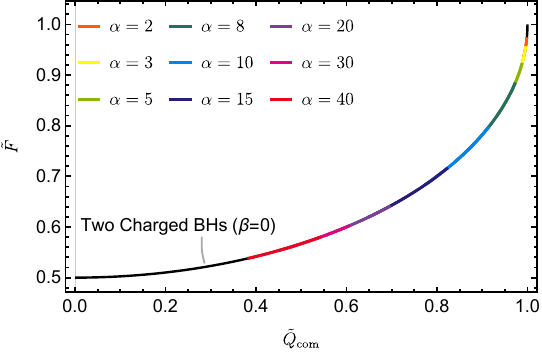}}
	\caption{
		\justifying
		\textbf{(a)}: The relationship between $\tilde{Q}_{com}$ and the reduced entropy $\tilde{S}$ for various coupling constants $ \alpha $.
		\textbf{(b)}: The relationship between $\tilde{Q}_{com}$ and the reduced entropy $\tilde{S}$ for various coupling constants $ \alpha $. The black line represents TCBHs, while the colored solid lines represent VRNBHs at different $\alpha$. } \label{fig:pqsF}
\end{figure}

Having explored the complex interplay between the electromagnetic and vector fields in this model, the significance of introducing the concept of the combined charge becomes evident. We now focus on examining how the thermodynamic properties of black holes in this combined charge framework differ from those in the single charge framework. By analyzing these thermodynamic properties, we can gain deeper insights into the stability and relative preferences of various black hole configurations.

\subsection{Thermodynamic Quantities}

For a vectorized black hole solution, it is crucial to examine thermodynamic quantities to analyze the preferences between VRNBHs, RNBHs, and TCBHs. By comparing parameters such as entropy and free energy, we can determine which configuration is thermodynamically favored.

If we consider the charge $ \tilde{Q} $ as the primary parameter, we encounter apparent thermodynamic inconsistencies. As evident from \cref{fig:sq}, the reduced entropy of VRNBHs is consistently greater than that of RNBHs, suggesting that VRNBHs should be favored according to the entropy criterion. However, the reduced free energy of VRNBHs is also greater than that of RNBHs, which seemingly contradicts the principle that lower free energy states are preferred (see \cref{fig:Fq}).

This apparent contradiction is resolved when we adopt the combined charge viewpoint. As illustrated in \cref{fig:pqsF}, we observe a remarkable phenomenon: while the reduced entropy of VRNBHs still exceeds that of TCBHs (represented by the black solid line), the reduced free energy of VRNBHs equals that of TCBHs. This perfect equivalence of free energies is a surprising result, given the distinct nature of these black holes. This alignment indicates that for a given combined charge, the system indeed favors VRNBHs in terms of thermodynamics, thereby resolving the apparent thermodynamic inconsistencies.

Moreover, our numerical calculations consistently support the above observations regarding the thermodynamics across various parameter ranges. Furthermore, this phenomenon of free energy can be demonstrated analytically.
From the definitions of free energy and the identity of horizon mass in \cref{eq:Feq,eq:Smarrrh}, along with the ADM mass ($M$) of both TCBHs and VRNBHs as given by \cref{eq:MeqPQ}, we obtain:
\begin{equation}
	\begin{aligned}
		F & =M-T_HS=M-\frac{1}{2}M_H=M-\frac{1}{2}\sqrt{M^{2}-\left(Q^{2}+P^{2}\right)} \\\implies\tilde{F}&=1-\frac{\sqrt{1-\left(\tilde{Q}^{2}+\tilde{P}^{2}\right)}}{2}
	\end{aligned}
\end{equation}
This equation clearly demonstrates that the reduced free energy $\tilde{F}$ depends on the combined charge $(\tilde{Q}^{2}+\tilde{P}^{2})$, not solely on $\tilde{Q}$, further supporting our combined charge perspective.

The introduction of the generalized coordinate not only resolves the apparent contradictions but also reveals new and intriguing features in the extended region (the region above the pseudo line in \cref{fig:qsF}). A particularly noteworthy phenomenon is observed in the relationship between $ \tilde{F} $ and $ \tilde{Q} $, as shown in \cref{fig:Fq}. In the extended region unveiled by our analysis, the free energy exhibits a non-monotonic behavior, suggesting a richer structure of thermodynamic stability than previously understood. This non-monotonicity indicates the presence of multiple branches of solutions. Such behavior is not captured in the BL coordinates and underscores the significance of our generalized coordinate approach.

In summary, the EMV model reveals that viewing thermodynamic quantities solely through the perspectives of charge $\tilde{Q}$ leads to apparent contradictions. This single-parameter perspective overlooks the critical influence of the vector field on black hole dynamics. The introduction of combined charge resolves these issues, offering a consistent and physically plausible framework for understanding VRNBH thermodynamics. Furthermore, the generalized coordinates enrich our understanding by unveiling complex thermodynamic structures, such as the non-monotonic behavior of free energy in extended regions. Our findings emphasize the necessity of considering both electromagnetic and vector contributions for an accurate assessment of thermodynamic stability. This approach ultimately provides a more comprehensive and physically grounded understanding of black hole behavior within the EMV model, highlighting the intricate interplay between various fields and their collective impact on black hole thermodynamics.

\subsection{Light Rings Radius}

\begin{figure}[htbp]
	\centering
	\subcaptionbox{\label{fig:rlrq}}
	{\includegraphics[width = 0.48\textwidth]{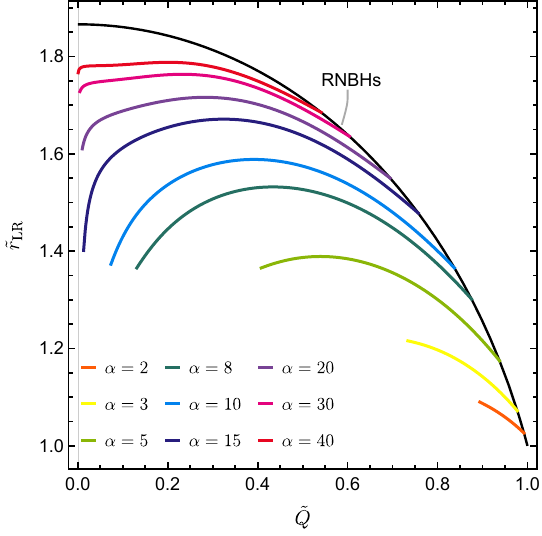}}
	\hfill
	\centering
	\subcaptionbox{\label{fig:rlrqp}}
	{\includegraphics[width = 0.47\textwidth]{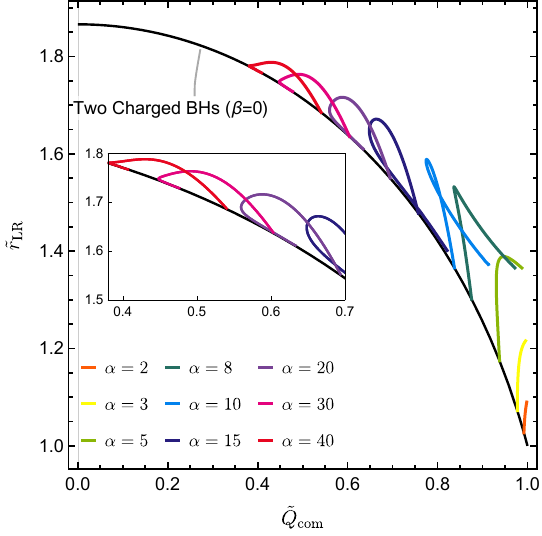}}
	\caption{
		\justifying
		\textbf{(a)}: Relationship between light ring radius $\tilde{r}_{LR}$ and reduced charge $q$ for VRNBHs with various coupling constants $\alpha$, compared to RNBHs (black line). \textbf{(b)}: Relationship between light ring radius $\tilde{r}_{LR}$ and combined charge $\tilde{Q}_{com}$ for VRNBHs with various $\alpha$, compared to TCBHs (black line). For $\alpha > 15$, the second branch of VRNBHs approaches or coincides with TCBHs at higher combined charge values.}
	\label{fig:rlr}
\end{figure}

As previously discussed, the introduction of the combined charge is crucial for understanding VRNBHs. The behavior of the light ring radius in VRNBHs reflects the influence of the vector field on the spacetime structure. Interestingly, two distinct perspectives on the light ring radius yield different insights.

\cref{fig:rlrq} illustrates the relationship between the light ring radius $\tilde{r}_{LR}$ and the reduced charge $\tilde{Q}$ for various coupling constants $\alpha$. The black line, representing $\tilde{r}_{LR}$ for RNBHs, serves as a baseline for comparison. For VRNBHs, we observe that $\tilde{r}_{LR}$ is consistently smaller than that of RNBHs across all values of $\alpha$. The curves show a consistent trend, clearly demonstrating the influence of $\tilde{Q}$ on the light ring radius. Notably, the difference in $\tilde{r}_{LR}$ between VRNBHs and RNBHs increases as $\tilde{Q}$ decreases.

In contrast, \cref{fig:rlrqp} presents the relationship between $\tilde{r}_{LR}$ and $\tilde{Q}_{com}$ for various $\alpha$ within the TCBHs framework. Here, the black line corresponding to TCBHs provides the baseline. Notably, in this framework, $\tilde{r}_{LR}$ of VRNBHs is consistently larger than that of TCBHs.

A particularly interesting feature emerges for larger $\alpha$ values ($\alpha >15 $), as shown in \cref{fig:rlrqp}. For these coupling constants, considering the second branch of VRNBHs discussed in \cref{sec:domain}, we observe that the light ring radius of this branch approach or even coincide with those of TCBHs as the combined charge increases. This convergence suggests a complex interplay between the vector field coupling strength and the combined charge in determining the spacetime geometry near the black hole.

These contrasting perspectives highlight the rich phenomenology of VRNBHs and underscore the importance of considering combined charge frameworks when analyzing their properties. The observed behaviors have potential implications for gravitational lensing, black hole shadows, and other observable phenomena in these exotic spacetimes.

\section{Discussion}\label{sec:4}

In this paper, we have presented a comprehensive analysis of spontaneous vectorization in the Einstein-Maxwell-Vector model. Our investigation has yielded several novel results on the vectorization.

First, by adopting a generalized coordinate, we eliminated the apparent divergences in the vector field near the event horizon that were present in previous studies using Boyer-Lindquist coordinates \cite{Oliveira:2020dru}. This generalized coordinate allowed us to extend the domain of existence for vectorized Reissner-Nordström black holes (VRNBHs) and provide a more complete picture of their properties. Our results demonstrate that the previously observed divergences were indeed coordinate artifacts rather than physical singularities, highlighting the importance of choosing appropriate coordinate systems when studying black hole theories.

Second, we introduced the concept of a combined charge $\tilde{Q}_{com}\equiv \sqrt{\tilde{Q}^2 + \tilde{P}^2}$, which incorporates both the reduced electric charge $\tilde{Q}$ and the reduced vector ``charge'' $\tilde{P}$. This approach was proven to be particularly insightful, revealing striking similarities between VRNBHs and two-charge black holes (TCBHs). The combined charge framework allowed us to resolve apparent thermodynamic inconsistencies that arise when considering only the electric charge. Our analysis shows that VRNBHs are thermodynamically favored over both RNBHs and TCBHs for a given combined charge.

Furthermore, our investigation of the light ring structure uncovered interesting behaviors that depend on the perspective taken. When viewed in terms of the electric charge $\tilde{Q}$, VRNBHs consistently exhibit smaller light ring radii compared to RNBHs. However, when analyzed using the combined charge framework, VRNBHs display larger light ring radii than TCBHs. This dual perspective enriches our understanding of how vectorization affects the spacetime geometry near black holes and could have significant implications for gravitational lensing and black hole shadow observations.

Morever, we identified a competition between the electromagnetic and vector fields, governed by the coupling constant $\alpha$. This competition manifests in the inverse correlation between $\tilde{Q}$ and $\tilde{P}$, and becomes more pronounced as $\alpha$ increases. Our analysis reveals that in the perturbative regime (small $\alpha$ or small $\tilde{P}$), the system closely resembles TCBHs, with deviations becoming more significant as $\alpha$ increases.

While our study has provided valuable insights into the EMV model, several open questions remain for future investigation. The stability of VRNBHs against perturbations needs to be rigorously analyzed to determine whether these solutions are physically realizable. Extending this work to rotating black holes could reveal even richer phenomenology, potentially offering more observationally relevant signatures. The dynamical process of spontaneous vectorization remains an intriguing area for further study, as it could shed light on the formation mechanisms of these exotic black holes. Additionally, exploring extensions of this model to include multiple vector fields or more complex coupling functions may uncover even more exotic black hole solutions. These future directions promise to deepen our understanding of black hole physics within the context of the Einstein-Maxwell theory with additional fields.

\section*{Acknowledgments}
We thank Bin Wu for helpful discussions. Peng Liu would like to thank Yun-Ha Zha for her kind encouragement during this work. This work is supported by the Natural Science Foundation of China under Grant No. 12475054, 12375048 and the Science and Technology Planning Project of Guangzhou (202201010655).


\appendix
\section{The Reissner-Nordström Black Hole in Generalized Coordinates}\label{sec:AA}
The Reissner-Nordström solution solves the Einstein-Maxwell field equation
\begin{equation}
	\begin{aligned}
		 & R_{ab}-\frac{1}{2}g_{ab}R=2\left({F_{a}}^{c}F_{bc}-\frac{1}{4}g_{ab}F_{cd}F^{cd}\right) \\
		 & \nabla _{a}F^{ab}=0
	\end{aligned}
\end{equation}
With the ansatz of \cref{eq:ansatz} the RN BH solution reads:
\begin{equation}
	\label{eq:hgRN}
	\begin{aligned}
		h & =\frac{r^2 (r+r_{H})^2}{\left(r(M+r)+r_{H}^2\right)^2} \\
		g & = \left(1+\frac{r_{H}}{r}\right)^{2}                   \\
	\end{aligned}
\end{equation}
together with the 4-potential
\begin{equation}
	\label{eq:AtRN}
	A_{t}=\frac{\left(r-r_{H}\right)^2 \sqrt{M^2-4r_{H}^2}}{\left(2 r_{H}+M\right) \left(r_{H}^2+r(M+r)\right)}
\end{equation}
where $ Q $ is the electric charge and we have been chosen gauge $A_{t}|_{r_{H}}=0$. The mass,charge and $r_{H}$ is related by:
\begin{equation}
	\label{eq:rhRN}
	r_{H} = \frac{1}{2}\sqrt{M^{2}-Q^{2}}=\frac{M}{2}\sqrt{1-q^{2}},\qquad q\equiv \frac{Q}{M}
\end{equation}
Note that the RN BHs in a generalized coordinate system
presented in \cref{eq:ansatz} can be obtained from the standard textbook Boyer-Lindquist
coordinates solution with the radial coordinate transformation
\begin{equation}
	\bar{r}_{BL}=r+M+\frac{M^{2}-Q^{2}}{4r}
\end{equation}

\section{Numerical Implementation}\label{sec:NI}
To begin, we substitute our proposed ansatz (\cref{eq:ansatz}) into the equations of motion (\cref{eq:space time,eq:vector field,eq:gauge field}). The resulting explicit forms are as follows:

\begin{align}
	 & \notag \partial _{r}^{2}\left[\log\left(\frac{h(r)}{g(r)}\right)\right]+\frac{1}{r}\partial _{r}\left[\log\left(\frac{h(r)^{2}}{g(r)^{2}}\right)\right]-\left\{\frac{1}{2}\partial _{r}\left[\log\left(\frac{h(r)}{g(r)}\right)\right]\right\}^{2}+\frac{B_t'(r)^2}{g_{tt}(r)} \\
	 & -\left(1+\frac{2\alpha B_t(r)^{2}}{g_{tt}(r)}\right)\mathcal{F}(r)=0,\label{eq:h2}                                                                                                                                                                                             \\
	 & \left(\partial _{r}\left[\log\left(\frac{h(r)}{r^{2}g(r)}\right)\right]\right)\frac{\mathcal{N}'(r)}{\mathcal{N}(r)}+\frac{h'(r)^2}{4 h(r)^2}-\frac{g'(r)^2}{4 g(r)^2}-\frac{g'(r)}{r\cdot g(r)}+\frac{B_t'(r)^{2}}{g_{tt}(r)}-\mathcal{F}(r)=0,\label{eq:g1}                  \\
	 & \partial _{r}\left(\frac{g'(r)}{g(r)}\right)+\frac{g'(r)}{r\cdot g(r)}-\frac{2 r_H}{r^2 \left(r-r_H\right)}+\frac{2h'(r)\mathcal{N}'(r)}{h(r)\mathcal{N}(r)}+\frac{h'(r)^{2}}{2h(r)^{2}}+\frac{2B_t'(r)^2}{g_{tt}(r)}-2\mathcal{F}(r)=0,\label{eq:g2}                          \\
	 & B_t''(r)-\left(\partial _{r}\left[\log\left(\frac{h(r)\mathcal{N}(r)}{r^{2}\sqrt{g(r)}}\right)\right]\right)B_t'(r)+ \alpha \mathcal{F}(r)B_t(r)=0,\label{eq:V2}                                                                                                               \\
	 & A_t''(r)-\left(\partial _{r}\left[\log\left(\frac{h(r)\mathcal{N}(r)}{r^{2}\sqrt{g(r)}}\right)\right]-\alpha \partial _{r}\left(\frac{B_t(r)^{2}}{g_{tt}(r)}\right)\right)A_t'(r)=0,\label{eq:A2}
\end{align}
where
\begin{equation*}
	g_{tt}(r)=-h(r)\mathcal{N}(r)^{2},\quad \mathcal{F}(r)=\frac{A_t'(r)^{2}}{h(r)\mathcal{N}(r)^{2}}\exp \left[{-\frac{\alpha {B_t(r)}^2}{h(r) \mathcal{N}(r)^{2}}}\right],\quad \mathcal{N}(r)=1-\frac{r_H}{r}
\end{equation*}

Our study now faces a set of highly nonlinear second-order ordinary differential equations (ODEs) involving the unknown functions $ h, g, A_t, $ and $ B_t $. Due to the complexity, we employ a pseudo-spectral method\footnotemark[4]
\footnotetext[4]{For technical details of pseudo-spectral methods in black hole physics, see \cite{Fernandes:2022gde}.} for numerical solutions. To achieve more precise numerical results, we perform further simplifications.

Notably, the Maxwell equation (\cref{eq:A2}) contains both first and second-order derivatives of $A_t(r)$, while the remaining four equations only involve the first derivative $A_t'(r)$, and not $A_t(r)$ itself. This characteristic allows us to integrate \cref{eq:A2} once, yielding an expression for $A_t'(r)$ which takes the form:
\begin{equation}
	A_t'(r)=\frac{Q(r-r_H)h(r)}{r^3 \sqrt{g(r)}} \exp \left[\frac{r^{2}\alpha B_t(r)^{2}}{(r-r_H)^{2}h(r)}\right] ,\label{eq:dAtdr}
\end{equation}
where $Q$ is the charge of the black hole. An additional simplification arises from the ability to express $A'_t$ in terms of $h$ and $B_t$. This allows us to eliminate $A_t$, solve for the other fields, and subsequently determine $A_t$ using these solutions.

On the other hand, upon careful examination, it becomes evident that \cref{eq:g1} and \cref{eq:g2} can be combined to eliminate the terms involving $ h(r) $ and $ A_t(r) $. This results in a single equation solely for $ g(r) $ as follows:
\begin{equation}
	\frac{g''(r)}{g(r)}+\frac{(3 r-{r_H}) g'(r)}{r (r-{r_H}) g(r)}-\frac{g'(r)^2}{2 g(r)^2}+\frac{2 {r_H}}{r^2 (r-{r_H})}=0.
	\label{eq:g22}
\end{equation}
By imposing boundary conditions that ensure regularity at the horizon ($g(r_H)=\operatorname{const}$) and asymptotic flatness at infinity ($g(\infty)=1$), we can derive an analytical solution for $g$.
\begin{equation}
	g(r)=\left(1+\frac{r_H}{r}\right)^2.
	\label{eq:g3}
\end{equation}

Substituting \cref{eq:dAtdr,eq:g3} into the remaining two equations (\cref{eq:h2,eq:V2}), we obtain the final ODEs that we need to solve numerically. The final forms are given by:
\begin{align}\label{eq:finaleom}
	 & \notag\frac{h''(r)}{h(r)}-\frac{3 h'(r)^2}{2 h(r)^2}
	+\frac{2 \left(r^2-r_H^2-r\cdot  r_H\right) h'(r)}{r h(r) \left(r^2-r_H^2\right)}
	+\frac{4 r_H^2}{r \left(r-r_H\right) \left(r_H+r\right)^2}                                                                                                                        \\
	 & +\frac{2 \alpha  Q^2 {B_t(r)}^2}{\left(r^2-r_H^2\right)^2}\exp \left[\frac{\alpha  r^2 {B_t(r)}^2}{h(r) \left(r-r_H\right)^2}\right]=0,                                        \\
	 & \notag B_t''(r)-\frac{h'(r) B_t'(r)}{h(r)}+\frac{\alpha  Q^2 h(r) B_t(r) }{r^2 \left(r_H+r\right)^2}\exp \left[\frac{\alpha  r^2 {B_t(r)}^2}{h(r) \left(r-r_H\right)^2}\right]
	\\&+\frac{2 \left(r^2-r_H^2-r\cdot r_H\right) B_t'(r)}{r\left(r^2-r_H^2\right)}=0.
\end{align}
To numerically solve the above ODEs, it is essential to impose appropriate boundary conditions. At spatial infinity, $r\to \infty$, we require the spacetime approach a Minkowski spacetime with vanishing vector fields:
\begin{align}
	 & h=1,\qquad B_t=0   .                                              \label{eq:bcin2}
\end{align}
On the other hand, at event horizon, $r\to r_H$, we require that all functions remain regular, ensuring that there are no singularities in the solution. Specifically, we set
\begin{equation}\label{eq:bcrh1}
	h-r_H \partial _{r}h=0,\qquad A_t=B_t=0.
\end{equation}
The condition on the vector field at the horizon arises from our requirement that $B_{a}B^{a}=(-B_t^{2})/(h \cdot \mathcal{N})$ remains regular. Since $\mathcal{N}  $ vanishes at the horizon, $B_t$ must be zero to satisfy this regularity condition. The conditions in \cref{eq:bcin2,eq:bcrh1} arise from their asymptotic behavior.
Near the horizon, all functions can be approximated by a power series as follows:
\begin{equation}
	\begin{aligned}
		 & h(r)=  h_0+\frac{h_0}{r_{H}}(r-r_{H})+\cdots   ,                 \\
		 & B_t(r)= b_2(r-r_{H})^{2}-\frac{b_2}{r_{H}}(r-r_{H})^{3}+\cdots .
	\end{aligned}
\end{equation}
At infinity, the requirement of asymptotic flatness necessitates that they take the following approximate forms:
\begin{equation}
	\begin{aligned}
		\label{eq:infexp}
		 & h(r)= 1-\frac{2(\sqrt{P^{2}+Q^{2}+4r_H^{2}}-r_{H})}{r}+\cdots ,\qquad
		 & B_t(r)= \frac{P}{r}+\cdots.
	\end{aligned}
\end{equation}
Additionally, for the boundary condition on $A_t$ in \cref{eq:bcrh1}, we have utilized the gauge freedom of the electromagnetic field. After obtaining the numerical solutions for $h(r), B_t(r)$, we use this boundary condition to solve $A_t(r)$ by integrating \cref{eq:dAtdr}.

To implement the pseudo-spectral method for solving \cref{eq:finaleom}, we introduce a compactified radial coordinate:
\begin{equation}
	z\equiv 1-\frac{2r_H}{r},
\end{equation}
which maps the range $r \in [r_H,\infty]$ to $z \in [-1,1]$. The corresponding boundary conditions will thus be transformed as follows:
\begin{equation}
	\begin{aligned}
		 & h=1,\quad                  & B_t & =0,\quad     & \text{for} \ z=1, &   \\
		 & h-2\partial _{z}h=0, \quad & A_t & =B_t=0,\quad & \text{for} \ z=-1 & .
	\end{aligned}
\end{equation}

In our numerical investigation, we utilized the Wolfram Engine, a free version of Mathematica, as our computational platform. To handle the nonlinearities in the equations of motion, we employed the Newton-Raphson iterative method combined with pseudospectral techniques to discretize the system. The stability of the iterative method is highly sensitive to the choice of initial guesses, so we used perturbative solutions as starting points. These perturbative solutions proved to be very effective in our numerics, as they are closely aligned with the expected physical behavior of the system. To systematically explore the parameter space, we adjusted the parameters $\alpha$ and $Q$ incrementally, allowing us to iteratively find nearby solutions. This stepwise adjustment was crucial for improving both the stability of the method and its efficiency, leading to faster convergence in our iterative process. This process continued until we reached parameter values where no further solutions could be obtained.

To ensure the validity of our solutions, we applied the Smarr relations (\cref{eq:smarr,eq:Smarrrh}) as a filtering mechanism. These thermodynamic relations serve as a consistency check for black hole solutions. We calculated the relevant quantities (mass, temperature, entropy, charge, and potential) from each obtained solution and substituted them into the Smarr relations. Solutions were considered valid if they satisfied \cref{eq:smarr} to within a tolerance of $10^{-4}$ and \cref{eq:Smarrrh} to within $10^{-8}$. This rigorous approach allowed us to identify and retain only the physically reasonable solutions within the VRNBHs parameter space.

\section{The Two Charged Black Holes} \label{sec:AB}

The action for two gauge fields is expressed as \cite{Zhang:2020znl}:
\begin{equation}
	S = \frac{1}{16\pi}\int d^4x \sqrt{-g} \left( R - F_{ab}F^{ab} - G_{ab}G^{ab} - 2\beta F_{ab}G^{ab}\right),\label{eq:actionTCBH}
\end{equation}
where $F = \mathrm{d}A$ and $G = \mathrm{d}B$ are the field strengths of the two gauge fields $A$ and $B$, respectively, and $\beta$ is the coupling constant. The Einstein equations derived from this action are given by:
\begin{equation}
	\label{eq:eomTCBH}
	\begin{aligned}
		R_{ab} - \frac{1}{2} R g_{ab} & = 2 T^{(A)}_{ab} + 2 T^{(B)}_{ab} + 4\beta T^{(AB)}_{ab}, \\
		\nabla_a F^{ab}               & = 0,                                                      \\
		\nabla_a G^{ab}               & = 0.
	\end{aligned}
\end{equation}
The energy-momentum tensors are defined as follows:
\begin{equation}
	\begin{aligned}
		T^{(A)}_{ab}  & = F_{ac} F_{b}{}^{c} - \frac{1}{4} g_{ab} F^2,                                             \\
		T^{(B)}_{ab}  & = G_{ac} G_{b}{}^{c} - \frac{1}{4} g_{ab} G^2,                                             \\
		T^{(AB)}_{ab} & = \frac{1}{2}(F_{ac} G_{b}{}^{c} + F_{bc} G_{a}{}^{c}) - \frac{1}{4} g_{ab} F_{cd} G^{cd}.
	\end{aligned}
\end{equation}

In BL coordinates, the two charged black hole solution is represented as:
\begin{equation}
	\begin{aligned}
		ds^{2}     & = -f(\bar{r})dt ^{2} + f(\bar{r})^{-1}dr^{2} + \bar{r}^{2}(d\theta ^{2} + \sin^{2}\theta d\phi ^{2}), \\
		f(\bar{r}) & = 1 - \frac{2M}{\bar{r}} + \frac{Q^{2} + P^{2} + 2\beta QP}{\bar{r}^{2}},                             \\
		A_{a}      & = \left( \Phi - \frac{Q}{\bar{r}}, 0, 0, 0 \right),                                                   \\
		B_{a}      & = \left( \Psi - \frac{P}{\bar{r}}, 0, 0, 0 \right).
	\end{aligned}
\end{equation}
Here, $Q$ and $P$ denote the charges associated with gauge fields $A_a$ and $B_a$, respectively, while $\Phi$ and $\Psi$ are the corresponding potentials. The horizon radius is defined as:
\[
	\bar{r}_{H} = M + \sqrt{M^{2} - (Q^{2} + P^{2} + 2\beta QP)}.
\]
From this solution, we can derive physical quantities:
\begin{equation}
	\begin{aligned}
		 & T_H = \frac{f'(\bar{r}_{H})}{4\pi}, \qquad S = \pi \bar{r}_{H}^{2},                   \\
		 & M = 2T_H S + Q \Phi + P \Psi + \beta (Q \Psi + P \Phi),                               \\
		 & M_H = 2T_H S, \quad \Phi = \frac{Q}{\bar{r}_{H}}, \quad \Psi = \frac{P}{\bar{r}_{H}}, \\
		 & M^{2}= Q^{2} + P^{2} + 2\beta QP + M_{H}^{2}.
	\end{aligned}
\end{equation}
In the limit as $\beta \to 0$, these relations simplify to:
\begin{equation}\label{eq:tcbhsmarr}
	\begin{aligned}
		 & M = 2T_H S + Q \Phi + P \Psi,      \\
		 & M^{2} = Q^{2} + P^{2} + M_{H}^{2}.
	\end{aligned}
\end{equation}
In this study, we concentrate on the scenario where \(\beta = 0\). Notably, in the EMV model, when the vector field is treated as a perturbation, the coupling term becomes negligible, causing its action to align with \cref{eq:actionTCBH} for \(\beta = 0\). However, beyond this perturbative limit, the two models differ significantly. By comparing the equations of motion from the two models (\cref{eq:vector2} and \cref{eq:eomTCBH}), a key distinction is that the vector field \(B_a\) in the EMV model possesses an effective mass, whereas this mass is absent in the other model.

\section{Proof: The Mass Stored in the Vector Field is Equal to Zero.}
\label{sec:AC}

Starting from the \cref{eq:vector field}, we contract both sides with $ B_{a} $ and integrate over the entire exterior spacetime of the black hole:
\begin{equation}
	\int_{\partial \Sigma _{t}} drd\theta d\varphi \sqrt{-g} B_{a}\nabla _{b}V^{ba}=\int_{\partial \Sigma _{t}} drd\theta d\varphi \sqrt{-g} U_{eff }^{2}B_{a}B^{a}
\end{equation}
The left side of the above equation:
\begin{equation}
	\begin{aligned}
		\int_{\partial \Sigma _{t}} drd\theta d\varphi \sqrt{-g}B_{a}\nabla _{b}V^{ba} & =\int_{\partial \Sigma _{t}} drd\theta d\varphi \sqrt{-g}\left[\nabla _{b}(B_{a}V^{ba})-V^{ba}\nabla _{b}B_{a}\right]
		\\&=-\int_{\partial \Sigma _{t}} drd\theta d\varphi \sqrt{-g}\left(\nabla _{b}B_{a}V^{ba}\right)
		\\&=\int_{\partial \Sigma _{t}} drd\theta d\varphi \sqrt{-g} U_{eff }^{2}B_{a}B^{a}\\
        \implies \nabla _{b}B_{a}V^{ba}&=-U_{eff }^{2}B_{a}B^{a}
	\end{aligned}
\end{equation}
The second equality holds because in spherical symmetry, the boundary conditions of  $B_a $ on the integrate surface are zero. And similarly:
\begin{equation*}
    \nabla _{a}B_{b}V^{ba}=U_{eff }^{2}B_{a}B^{a}
\end{equation*}
combine these two equations, we have:
\begin{equation}
\begin{aligned}
        V_{ab}V^{ab}+2U_{eff }^{2}B_{a}B^{a}&=0\\
      \implies  V_{rt}V^{rt}+U_{eff }^{2}B_{t}B^{t}&=0
\end{aligned}
\end{equation}
Moreover, considering \cref{eq:Mv}:
\begin{equation}
	\begin{aligned}
		M_{V} & \equiv \int_{\partial \Sigma_t}{dS}^{a}\left(2T_{ab}^{V}\xi^{b}-T^{V}\xi_{a}\right)=-\int_{\partial \Sigma _{t}} drd\theta d\varphi \sqrt{-g}\left(2{T^{t}}_{t}^{V}-T^{V}\right)
		\\&=-\int_{\partial \Sigma _{t}} drd\theta d\varphi \sqrt{-g}\left(V_{rt}V^{rt}+U_{ef f}^{2}B_{t}B^{t}\right)
		\\&=0
	\end{aligned}
\end{equation}
Therefore, the mass stored in the vector field is equal to zero.

\nocite{*}

\end{document}